Flat-Band Generation in InAs/GaSb Quantum Wells
through Vertically Engineered Heterostructures

Zachery A. Enderson[1,2,3,*†], Jiyuan Fang[4,†], Wei-Chen Wang[4], Li Xiang[5], Mykhaylo Ozerov[5], Dmitry Smirnov[5], Zhigang Jiang[4,*], Samuel D. Hawkins[6], Aaron J. Muhowski[6], John F. Klem[6], and Wei Pan[2,*]

[1] Oak Ridge Institute for Science and Education IC Postdoctoral Fellowship 37831, USA
[2] Sandia National Laboratories, Livermore, CA 94551, USA
[3] Institute for Matter and Systems, Georgia Institute of Technology, Atlanta, GA 30332, USA
[4] School of Physics, Georgia Institute of Technology, Atlanta, GA 30332, USA
[5] National High Magnetic Field Laboratories, Tallahassee, FL 32310, USA
[6] Sandia National Laboratories, Albuquerque, NM 87185, USA

*Corresponding author. Email: zenderson3@gatech.edu (ZAE);
zhigang.jiang@physics.gatech.edu (ZJ); wpan@sandia.gov (WP)
†These authors contributed equally to this work.

**Abstract**
Quantum materials constitute a novel category of substances wherein quantum effects and electron-electron (e-e) interactions give rise to unforeseen phenomena on a macroscopic scale. Of particular interest within the realm of quantum materials are flat bands, which promote heavy conduction electrons and enhance e-e correlation effects. While the engineering of such flat bands has been demonstrated in graphene and two-dimensional transition metal dichalcogenides moiré superlattices and in lithography defined semiconductor moiré superlattices, conventional tear-and -stack fabrication methods face challenges due to inevitable twist-angle disorder, strain, and relaxation effects, leading to issues with reproducibility and scalability. We explore the creation and modification of flat bands through vertically engineered III-V semiconductor heterostructures, without the need for twisting. These artificial quantum materials offer a reproducible and scalable means for producing high-quality flat-band materials via molecular beam epitaxy growth. Our investigation includes magnetotransport and infrared magneto-spectroscopy studies of quad-layer InAs/GaSb quantum wells, accompanied by k·p band structure calculations, which illustrate the flattening of bands in vertically designed heterostructures.

## I. INTRODUCTION

A full illustration of the electronic and magnetic properties of any material is, inherently, rooted in quantum mechanics but, for many materials, the quantum description becomes secondary on the macroscopic scale. Quantum materials represent an emerging class of materials in which quantum effects and electron-electron (e-e) interactions give rise to macroscopic phenomena [1]. Flat, nearly dispersionless, low-energy bands are especially intriguing within the realm of quantum materials because they can foster the presence of heavy conduction electrons and facilitate pronounced e-e correlation effects. The heightened e-e correlations can lead to superconductivity mediated by purely Coulombic forces – rather than phonon-mediated interactions – which is an enticing property for potential room temperature superconductors [2-5]. These flat bands have been demonstrated in lateral superlattices in two-dimensional (2D) materials and in compound semiconductor quantum wells [6], that is, the so-called moiré superlattice materials [5, 7-13]. But, conventional tear-and-stack fabrication methods have been troubled by challenges due to twist-angle disorder [14, 15], strain [16-18], and relaxation effects [19-21], leading to issues with reproducibility and scalability.

In this Article, we present an alternative method for generating flat-band systems through vertically engineered InAs/GaSb heterostructures, without the need for twisting. Due to the unique InAs/GaSb band alignment, different band topologies can be achieved simply by tuning the heterostructure parameters, such as the quantum well thickness [22, 23]. In bulk material form, the top of the valence band of GaSb is 0.143 eV higher than the bottom of the conduction band of InAs. Due to quantum confinement in the heterostructure, the alignment of E1 (the lowest electron subband in the conduction band) and H1 (the highest hole subband in the valence band) can be tuned by varying the well thicknesses. Consequently, the quantum well stack can be in any of three different regimes: a normal band (conduction band edge above valence band edge), a zero gap band (conduction band edge touching valence band edge), or an inverted band (conduction band edge below valence band edge) [24]. In the inverted regime, the electron and hole subbands hybridize and invert, opening a gap. As Fig. 1 displays, the resulting shape of the inverted bands depends on the degree of hybridization and, when carefully optimized, can result in flat bands. This work investigates how the flat-band regime can be enhanced by utilizing a quad-layer InAs/GaSb quantum well stack to increase the flattened region about the Γ-point – flatter than in any InAs/GaSb quantum well bilayers.

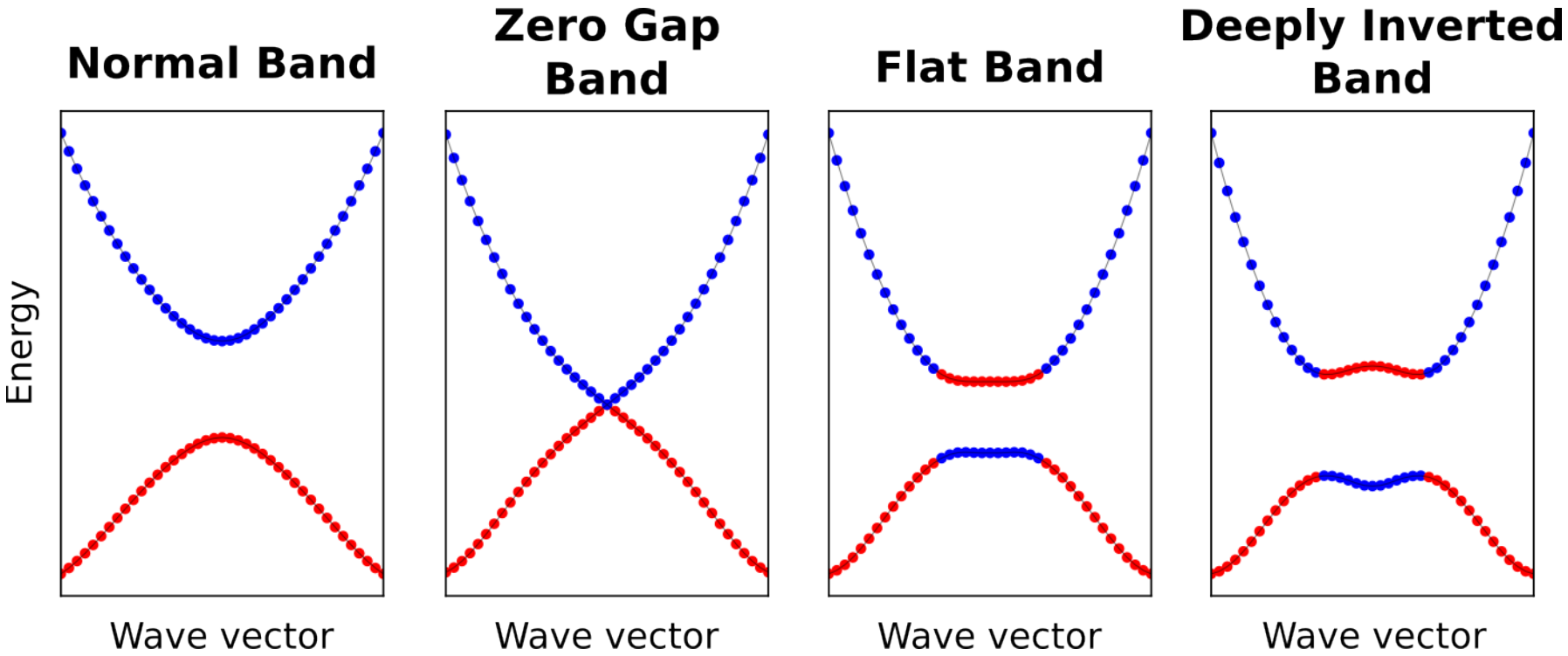

**Fig. 1. Effect of increasing band inversion.** Diagram of the band flattening process using k·p calculated band structures for varying quad-layer InAs/GaSb quantum well parameters.

## II. METHODS

### Computational methods

We adopt an eight-band k·p model to theoretically investigate the low-energy band structure and Landau levels of quad-layer InAs/GaSb quantum wells. The band structure is calculated using an adapted Kane model for zinc-blende semiconductors, which has been shown to be effective for systems with strong electron-hole hybridization [25]. The calculation also incorporates the effects of Zeeman splitting [26], strain induced by lattice mismatch [27], and uses Burt-Foreman envelope function theory to account for material variation across the heterogeneous structure [28-32]. The material parameters of InAs and GaSb used were obtained from the literature [33], which prove to be successful in explaining the results of our prior works [24, 34-37]. Additional details can be found in the Supplemental Material.

## III. EXPERIMENTAL DETAILS

### A. Material synthesis

High mobility, near surface quad-layer InAs/GaSb quantum well heterostructures were grown via molecular beam epitaxy (MBE) on a GaSb(001) substrate and situated between Al-containing barrier layers (Fig. 2B). We vary the thickness of each quantum well layer to tune the band structure of the material, guided by k·p calculations. Figure 2A shows the k·p calculated zero-field band structure for the uppermost valence bands and low-energy conduction bands for our chosen quad-layer sample – the 6,9,9,8-nm structure shown in Fig. 2B. This heterostructure was chosen to maximize the flat-band region that is generated around the Γ-point.

### B. Magnetotransport

Samples used for magnetotransport measurements were mounted on a ceramic chip carrier with gold contacts attached with indium solder. The measurements were performed at the National High Magnetic Field Laboratory (NHMFL) using an 18 T superconducting magnet equipped with a $^{3}$He system capable of achieving a base temperature of 300 mK. The measurements were performed using standard low frequency (< 20 Hz) lock-in techniques using a drive current of 100 nA.

### C. Far-infrared magneto-spectroscopy

Far-infrared magneto-spectroscopy (FIRMS) measurements were also performed at the NHFML using a 17.5 T superconducting magnet at 5 K. Spectroscopic measurements were made in the Faraday transmission configuration. The infrared (IR) spectrometer (Bruker VERTEX 80v) uses a mylar multilayer beam splitter for far-IR measurements and transmits the unpolarized IR radiation from a mercury lamp source to the sample via evacuated light pipes. The incident light has a beam diameter of ≈ 3 mm, and the intensity of the transmitted light is detected with a composite Si bolometer.

## IV. Results & Discussion

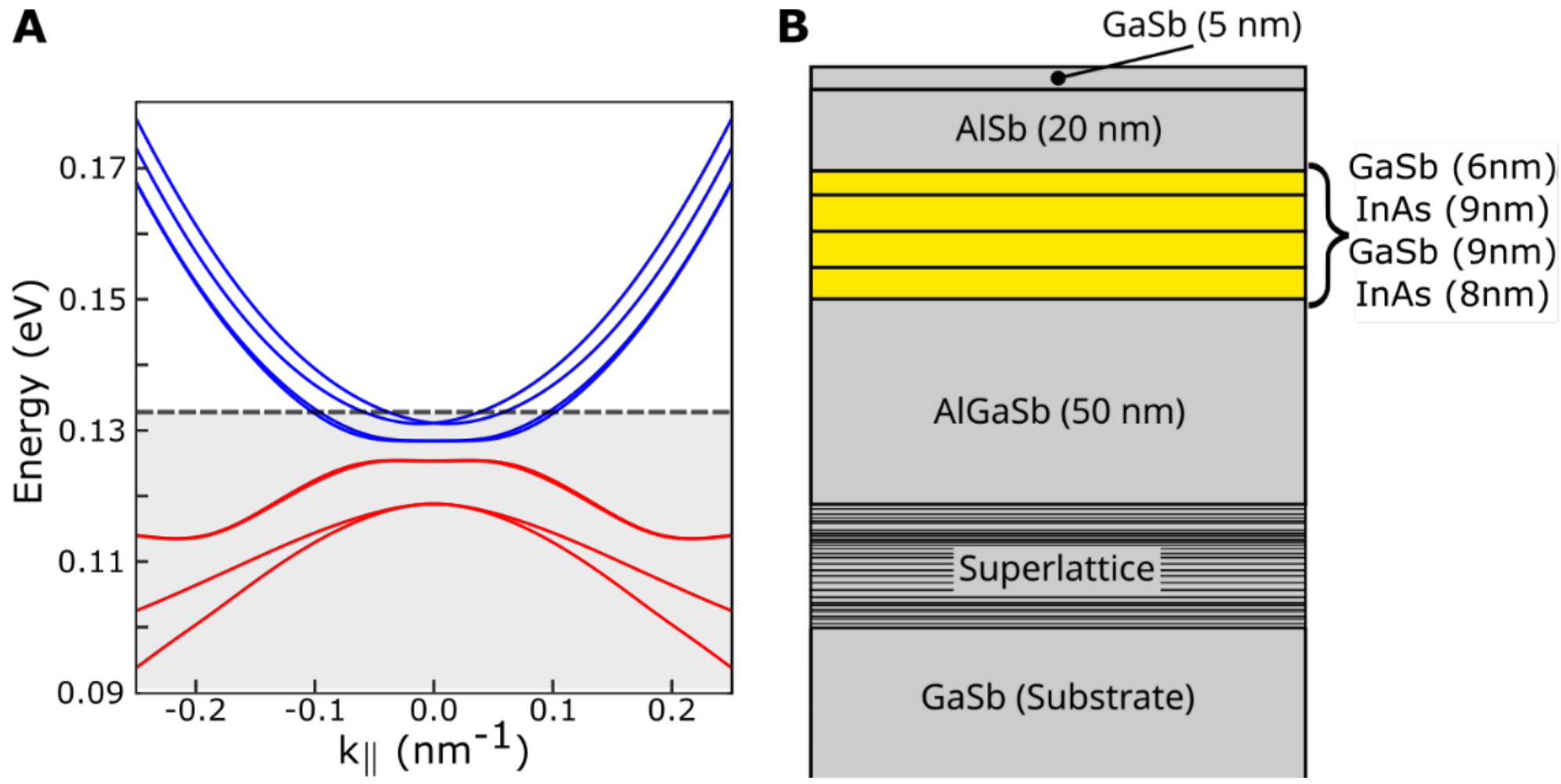

**Fig. 2. Sample structure and calculated band structure.** (**A**) k·p calculated zero-field band structure for the featured quad-layer InAs/GaSb quantum well sample. The conduction bands are in blue and valence bands in red. The dashed gray line indicates the experimental Fermi energy. (**B**) Structure diagram for the quad-layer InAs/GaSb sample with the quantum well layers depicted in yellow.

## A. Magnetotransport characterization

Initial magnetotransport characterization of a 6,9,9,8-nm quad-layer InAs/GaSb quantum well sample was performed at $T \approx 390$ mK. The magnetoresistance, $R_{xx}$, and the Hall resistance, $R_{xy}$, are displayed in Fig. 3. Well-defined Shubnikov-de Haas (SdH) oscillations and the integer quantum Hall effect are seen starting around 0.5 T. There is good agreement between the carrier densities calculated from the spacing of the SdH oscillations in $R_{xx}$ and the low magnetic field dependence of $R_{xy}$ giving a total carrier density (due to unintentional doping) of $n_e = 1.9 \times 10^{11}$ cm$^{-2}$. The sample had a Hall-bar geometry giving a total Hall mobility of $\mu = 7.1 \times 10^4$ cm$^2$V$^{-1}$s$^{-1}$.

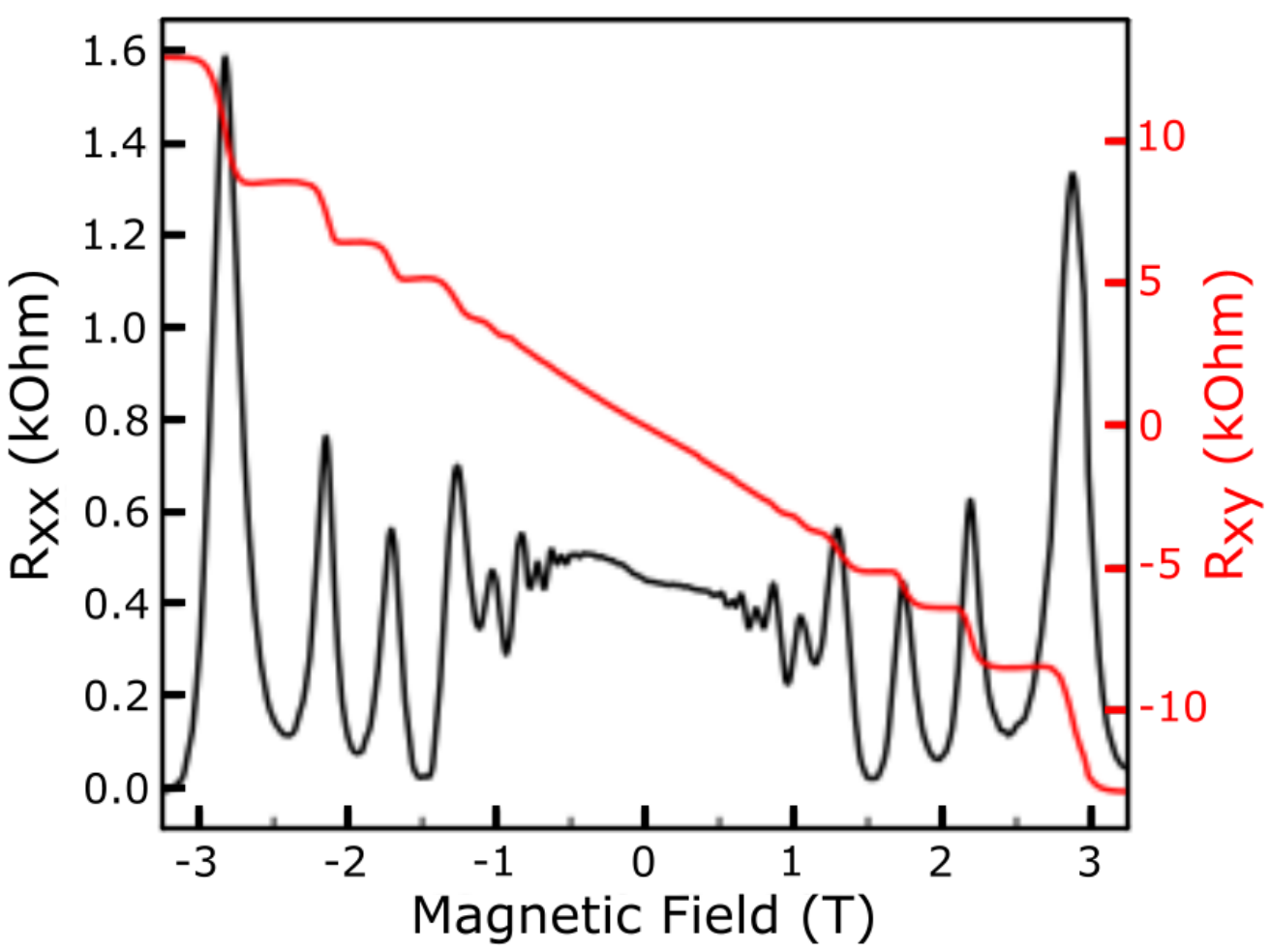

**Fig. 3. Magnetotransport sample characterization.** Magnetotransport measurements of a 6,9,9,8-nm quad-layer InAs/GaSb quantum well sample taken at ≈ 390 mK. The black curve (left axis) is the magnetoresistance $R_{xx}$ and the red curve (right axis) is the Hall resistance $R_{xy}$.

The conduction bands of the eight-band, zero-field k·p model are equated to a two-band model, in which the average of the two higher-energy conduction bands in Fig. 2A is defined as CB2 and the average of the lower-energy conduction bands is defined as CB1. Within this framework, we can fit the low-field experimental data using the two-band expression of the Hall conductivity [38],

$$\sigma_{xy} = -\frac{R_{xy}}{R_{xy}^2+W^2R_{xx}^2/L^2} = eB\left(-\frac{n_{\mathrm{CB1}}\mu_{\mathrm{CB1}}^2}{1+\mu_{\mathrm{CB1}}^2B^2} - \frac{n_{\mathrm{CB2}}\mu_{\mathrm{CB2}}^2}{1+\mu_{\mathrm{CB2}}^2B^2}\right), \tag{1}$$

where $W$ and $L$ are the width and length of the Hall-bar channel, $e$ is the electron charge, and $B$ is the out-of-plane magnetic field. The fitting is guided by our k·p band structure calculation of Fig. 2A, i.e., using the Fermi energy, $E_F$, as a fitting parameter, instead of $n_{\mathrm{CB1}}$ and $n_{\mathrm{CB2}}$. The best fit to data reads $E_F = 0.133$ eV (indicated by the dashed grey line in Fig. 2A), corresponding to $n_{\mathrm{CB1}} = 1.53 \times 10^{11}$ cm$^{-2}$ and $n_{\mathrm{CB1}} = 0.41 \times 10^{11}$ cm$^{-2}$. The k·p calculated effective mass, $m^*$, of each conduction band can then be obtained from

$$m^* = \hbar^2\left(\left.\frac{d^2E}{dk_{\parallel}{}^2}\right|_{E_F}\right)^{-1}, \tag{2}$$

where $\hbar$ is the reduced Planck's constant. As a result, $m_{\mathrm{CB1}}^* = 0.049\, m_e$ and $m_{\mathrm{CB2}}^* = 0.059\, m_e$, where $m_e$ is the free electron mass. We note that these calculated electron effective masses are significantly higher than that reported for InAs ($m^* = 0.023\, m_e$) [39] indicating a flatter band even at this Fermi energy. These values will be used as a method of comparison with our experimentally determined effective mass.

**B. Temperature dependent SdH oscillations**

Using the phenomenological model presented by Knobel *et al.* [40], we can directly compare the magnetotransport data with the k·p model in an external magnetic field as shown in Fig. 4. To achieve this agreement, the model has a carrier density of $n_e = 1.8 \times 10^{11}$ cm$^{-2}$, a conduction band offset of 3.8 meV (the energy offset between the conduction bands at the Γ-point), a disorder-induced broadening factor of $\gamma = 1.2\sqrt{B}$ meV, and a broadening factor of $\gamma_{\mathrm{ext}} = 0.25$ meV for the extended (delocalized) states. Here, the band offset accounts for the electric-field-induced band gap closing due to finite doping as the k.p band struture was origially calculated for the intrinsic case [34]. The slightly reduced value of the carrier density, compared to the experiment, is due to the magnitude of the broadening factor $\gamma$. $\gamma_{\mathrm{ext}} = 0.25$ meV corresponds to a transport relaxation time of $\tau \sim \hbar/\gamma_{\mathrm{ext}} = 2.6$ ps, consistent with the value of 2.0 ps deduced from the (low-field) carrier mobility in Eq. 1. In Fig. 4, the experimental and k·p results show the same Landau level occupancy which is especially clear at higher magnetic fields (such as at filling factors $\nu = 3, 4, 5, 7, 8$ ...). This is particularly evident by the absence of SdH oscillations and quantum Hall plateaus corresponding to $\nu = 6$ and $\nu = 11$ in the experiment which is precisely reproduced in the calculated model and found to be due to Landau level crossings between CB1 and CB2 at $B \approx 1.25$ T and 0.62 T, respectively (see supplemental Landau fan diagram, Figs. S1,S2).

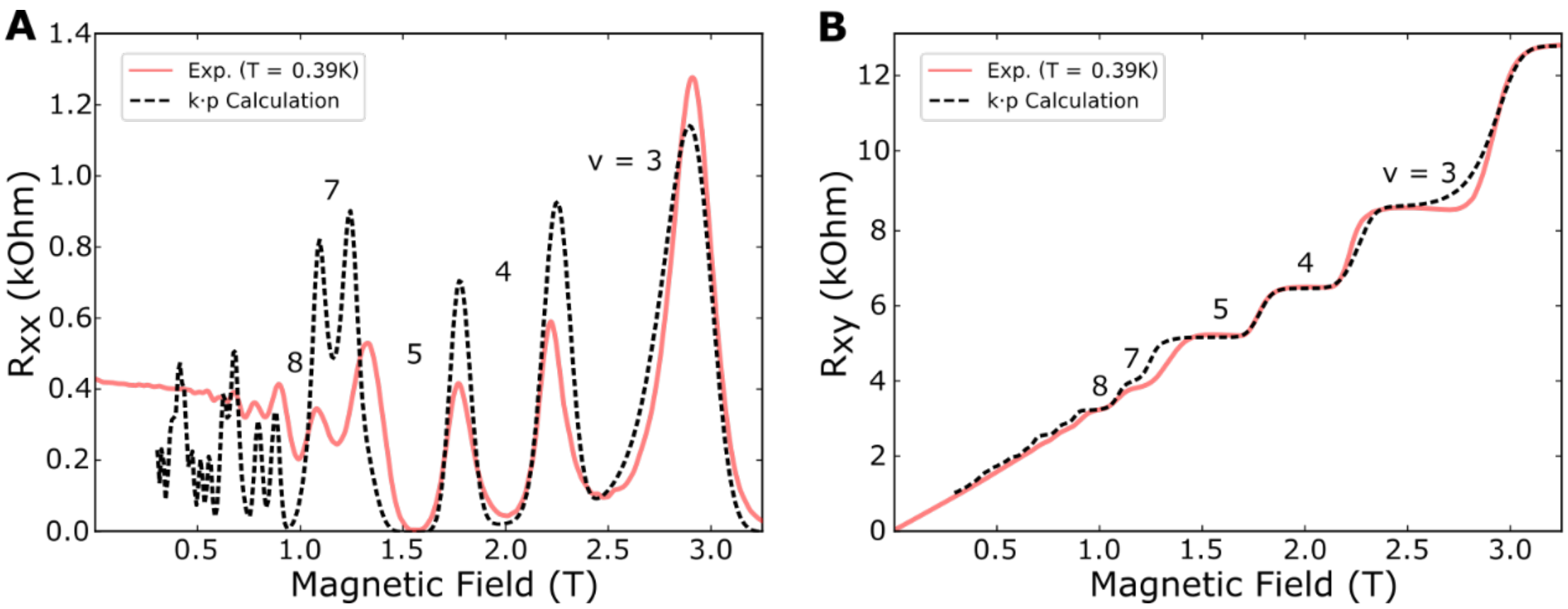

**Fig. 4. Experimental and theoretical magnetotransport data.** Comparison between experimental magnetotransport measurements (solid red lines) and k·p calculated values (black dashed lines) for (**A**) magnetoresistance $R_{xx}$ and (**B**) Hall resistance $R_{xy}$.

To experimentally determine the effective mass $m^*$ of the quad-layer system, we investigate the temperature dependence of the SdH oscillation amplitude of $R_{xx}$, which is given by the thermal dampening term in the Lifshitz-Kosevich (L-K) formula [41]

$$\Delta R_{xx} \propto \frac{\alpha T}{\sinh(\alpha T)} e^{-\alpha T_D}, \tag{3}$$

with

$$\alpha = \frac{2\pi^2 k_B}{\hbar\omega_c},$$

where $k_B$ is the Boltzmann constant, $T_D$ is the Dingle temperature, and $\omega_c$ is the cyclotron resonance frequency. An effective mass is found by fitting the SdH oscillation amplitude $\Delta R_{xx}$ with Eq. 3 to determine $\omega_c = eB/m^*$ and then the corresponding $m^*$ .

The data was acquired by sweeping the magnetic field between 0 and -3.75 T at a constant temperature within the range 0.75 to 10.75 K. A plot of this data is shown in the upper inset of Fig. 5, with a 2nd-order polynomial background removed. We determine the amplitude for each valley associated with an integer Landau level filling factor at each temperature (e.g., see the lower inset of Fig. 5) and fit the data with the L-K formula above (see Supplemental Material for details). The data and fit for $\nu = 5$ ($B \approx -1.53$ T) are shown in the main panel of Fig. 5. This procedure is repeated for all distinguishable filling factors, and the full set of results is featured in Table 1. The Dingle temperature could vary during the fit and is found to have an average value of $T_D = 3.2 \pm 0.2$ K. We note that the extracted effective mass at $\nu = 3$ ($B \approx -2.52$ T) is much higher than the average, presumably due to the crossing of the lowest Landau level of CB2 (which originates from the bottom of the band with a heavier mass) with a Landau level of CB1, as shown in the calculated Landau fan diagram in supplemental Figs. S1,S2. Averaging the effective masses measured with $\nu \geq 4$ gives $m^* = 0.053\, m_e$. The average effective mass calculated from the SdH oscillation amplitudes of individual filling factors agrees with that calculated using a separate Fourier transform analysis method on the same experimental data set using the oscillations associated with $\nu \geq 4$ (see supplemental Figs. S5,S6).

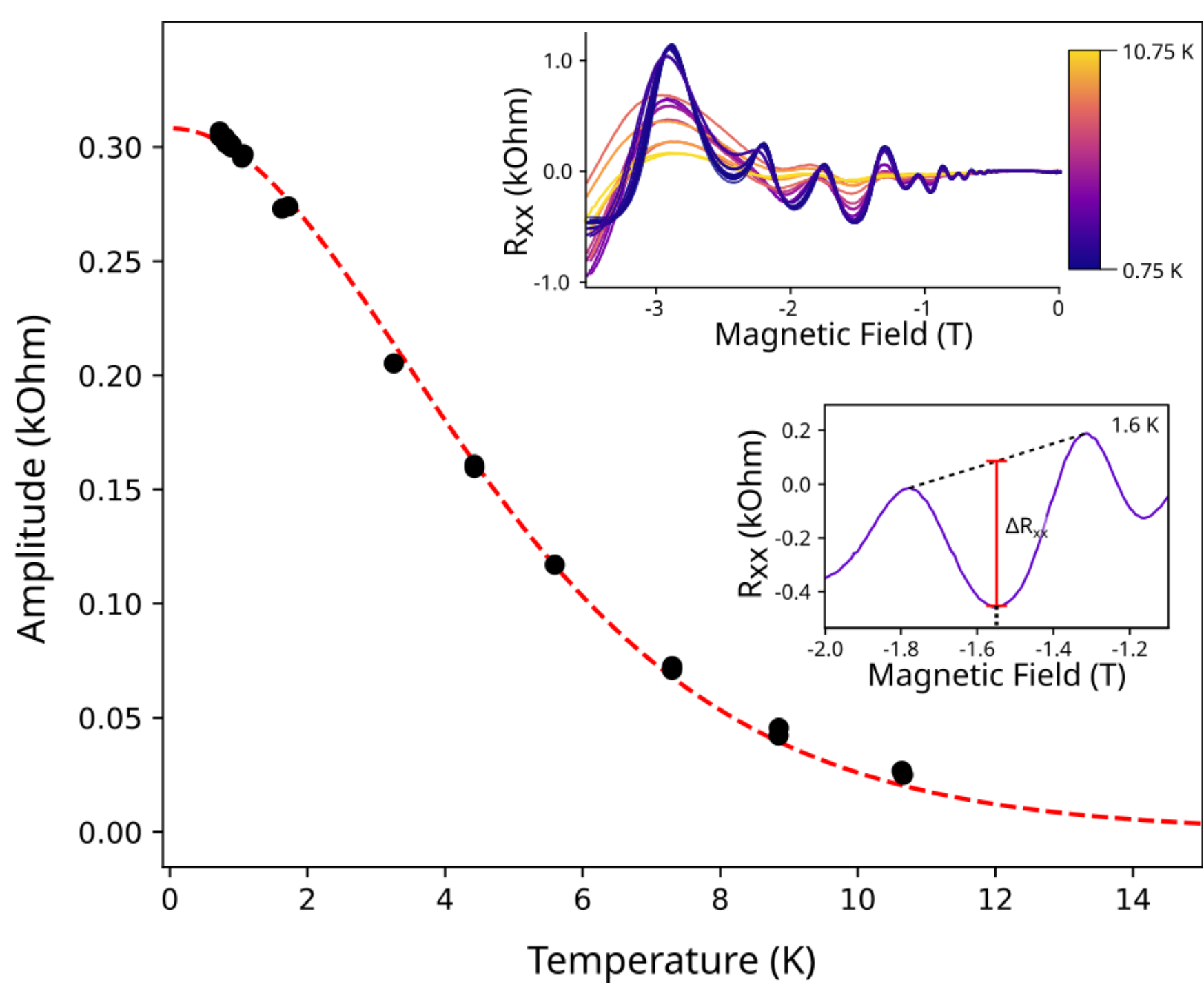

**Fig. 5. Effective mass from temperature dependent SdH oscillation amplitudes.** SdH oscillation amplitude dependence for filling factor $\nu = 5$ at $B \approx -1.53$ T. Black dots indicate experimental data points of the oscillation amplitude $\Delta R_{xx}$ found at various temperatures between 0.75 and 10.75 K. Dashed red line is from fitting with Eq. 3. Analysis of this SdH oscillation period has found an effective mass, $m^* = 0.048\, m_e$. (Upper inset) Experimental $R_{xx}$ curves of the temperature dependent SdH oscillation analysis presented. Each curve has had a 2nd-order polynomial background removed. (Lower inset) Illustration of the process extracting the amplitude of SdH oscillation at $\nu = 5$ (data taken at $T = 0.75$ K).

| **Filling Factor, $\nu$** | **3** | **4** | **5** | **7** | **8** | **9** | **10** | **12** |
|---|---|---|---|---|---|---|---|---|
| ***B*-Field (T)** | -2.52 | -2.03 | -1.53 | -1.16 | -0.95 | -0.80 | -0.70 | -0.62 |
| **$m^*$ ($m_e$)** | 0.243 | 0.075 | 0.048 | 0.057 | 0.047 | 0.056 | 0.044 | 0.044 |
| **Variance ($m_e$)** | 0.010 | 0.007 | 0.002 | 0.001 | 0.001 | 0.002 | 0.002 | 0.007 |

**Table 1.** Effective mass values (in units of free electron mass, $m_e$) from temperature dependent SdH oscillation amplitudes measured at the minima of the oscillations given by the specified magnetic field values and the corresponding filling factors.

## C. FIRMS measurements

We use FIRMS as an additional experimental method to confirm the modified band structure. A $1 \times 1$ cm$^2$ quad-layer InAs/GaSb quantum well sample was studied in the Faraday transmission configuration at $T \approx 5$ K. For the determination of the effective mass, we are only interested in the cyclotron resonance mode (intraband transition) which resides in the far-IR regime (10–650 cm$^{-1}$) and dominates within magnetic fields $< 9.0$ T. Figure 6A shows the normalized transmission, $T(B)/T_{ave}$, through our sample, with $T_{ave}$ being the average spectrum of all magnetic fields. The spectra $T(B)$ were taken at constant magnetic field with 0.5 T increments between scans. More measurement and analysis details can be found in the Supplemental Material.

In addition to the theoretical zero-field band structure calculation of the effective mass (Eq. 2), we expanded the k·p model to calculate the Landau levels in a magnetic field and the resulting absorption spectrum [34]. This calculated absorption spectrum is shown in Fig. 6B. The allowed inter-Landau-level optical transitions are described by Fermi's golden rule with respect to the electric dipole transition (the leading term of the incident light perturbating Hamiltonian). It also considers electron state occupancy at the experimental temperature of 5 K. This model uses a carrier density of $n_e = 1.9 \times 10^{11}$ $cm^{-2}$, consistent with the magnetotransport measurements above.

In the low-energy regime, the absorption is dominated by the cyclotron resonance and is found experimentally to have a slope of 0.059 $T/cm^{-1}$, giving an $m^* = 0.055\, m_e$. This value is in excellent agreement with that of the zero-field band structure and the average effective mass for $\nu \geq 4$ in the temperature dependent SdH oscillation analysis. Furthermore, the fit to the cyclotron resonance mode (red dashed line in Fig. 6A) is in good agreement with the low-energy calculated absorption spectrum which exhibits the same onset $B$-field and slope. The slight splitting of the cyclotron resonance mode observed in the experimental results around $> 90$ $cm^{-1}$ may arise from an interband Landau level transition that is not included in our k.p calculation due to the unknown band offset (band gap) resulting from the finite doping effect mentioned above.

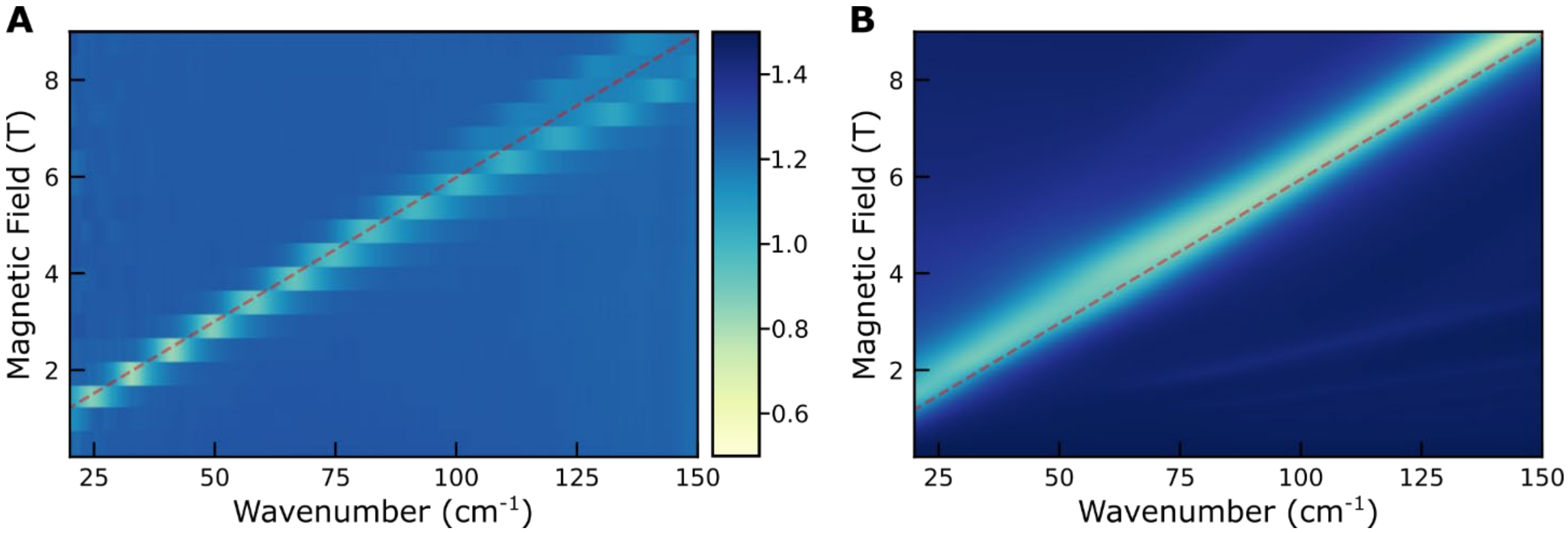

**Fig. 6. Cyclotron resonance mode via FIRMS measurement and k·p calculation.** (**A**) False color FIRMS results of the quad-layer InAs/GaSb sample displaying the magnetic field dependence of the cyclotron resonance mode. The color indicates the normalized transmission, $T(B)/T_{ave}$, through the sample at a given incident wavenumber (horizontal axis) and external magnetic field (vertical axis). The red dashed line is the linear fit used to calculate the effective mass, $m^* = 0.055\, m_e$. (**B**) k·p calculated optical absorption false color plot over the same magnetic field and energy range as (A). The red dashed line is the linear fit from the experimental results to aid comparison.

## D. Discussion

Following a previously identified relationship between the thicknesses of InAs and GaSb layers in a double quantum well heterostructure and the magnitude of valence/conduction band inversion [34], we chose to fabricate a quad-layer InAs/GaSb system that we predicted using k·p calculations to have flat bands near the Γ-point. The 6,9,9,8-nm thick GaSb/InAs/GaSb/InAs quantum well heterostructure grown by MBE appeared to be of high quality as shown through low-temperature magnetotransport measurements by its high mobility ($\approx 10^5$ $cm^2V^{-1}s^{-1}$). As a result, well-developed SdH oscillations and quantum Hall effect were observed. The zero-field k·p band structure calculation predicts an effective mass of $\approx 0.054\, m_e$ at the experimental determined charge density (averaging the values of CB1 and CB2). Temperature dependent SdH oscillation measurements extracted an effective mass of $0.053\, m_e$ in low magnetic fields. As an additional experimental approach, we performed FIRMS measurements on the sample to map the energy of the cyclotron resonance mode as a function of external magnetic field from which we found an effective mass of $0.055\, m_e$. We compared this to theoretical calculations of the cyclotron absorption energies and found them to agree with the experimental data.

## V. SUMMARY

The results presented in this work demonstrate that our k·p calculations accurately and consistently reproduce the experimental results for the quad-layer InAs/GaSb quantum well sample when using the experimentally measured carrier density, thereby supporting the validity of the model. This model shows that there exists a regime of structural parameters where quad-layer InAs/GaSb samples exhibit a high degree of band flattening around the Γ-point in the lowest (highest) energy conduction (valence) band, due to band inversion. Even with an unintentional doping of $n_e = 1.9 \times 10^{11}$ cm$^{-2}$, the effective mass is found to be $\approx 0.054\, m_e$, significantly higher than that reported for InAs ($m^* = 0.023\, m_e$) [39]. Such III-V heterostructure samples may represent one specific example in a family of materials that can be used to explore emerging flat-band physics and practical applications.

## ACKNOWLEDGEMENTS

This research is supported by an appointment to the Intelligence Community Postdoctoral Research Fellowship Program (ZAE) at Sandia National Laboratory (SNL) administered by Oak Ridge Institute for Science and Education through an interagency agreement between the U.S. Department of Energy (DOE) and the Office of the Director of National Intelligence. ZAE also acknowledges support from the Georgia Tech Institute for Matter and Systems, a member of the National Nanotechnology Coordinated Infrastructure, which is supported by the National Science Foundation (NSF, ECCS-2025462). ZJ and DS acknowledge support from the DOE (DE-FG02-07ER46451), and ZJ acknowledges support from the Sandia Academic Alliance Program. Measurements at NHMFL are supported by the NSF (DMR-2128556) and the State of Florida. WP acknowledges support from the DOE, Office of Science, Basic Energy Sciences, as part of the Microelectronics Energy Efficiency Research Center for Advanced Technologies (MEERCAT), a Microelectronics Science Research Center (MSRC). The work at SNL is supported by the Laboratory Directed Research and Development program and a user project at the Center for Integrated Nanotechnologies, an Office of Science User Facility operated for the DOE Office of Science. SNL is a multimission laboratory managed and operated by National Technology and Engineering Solutions of Sandia, LLC., a wholly owned subsidiary of Honeywell International, Inc., for the DOE's National Nuclear Security Administration (DE-NA-0003525). This paper describes objective technical results and analysis. Any subjective views or opinions that might be expressed in the paper do not necessarily represent the views of the U.S. DOE or the U.S. Government.

## DATA AVAILABILITY

The data that support the findings of this article are not publicly available. The data are available from the authors upon reasonable request.

## References:

[1] B. Keimer, and J.E. Moore, *"The physics of quantum materials,"* Nature Physics **13**, 1045-1055 (2017).
[2] M.R. Norman, *"The Challenge of Unconventional Superconductivity,"* Science **332**, 196-200 (2011).

[3] N.B. Kopnin, T.T. Heikkilä, and G.E. Volovik, *"High-temperature surface superconductivity in topological flat-band systems,"* Physical Review B **83**, 220503 (2011).
[4] H. Aoki, *"Theoretical Possibilities for Flat Band Superconductivity,"* Journal of Superconductivity and Novel Magnetism **33**, 2341-2346 (2020).
[5] L. Balents, C.R. Dean, D.K. Efetov, and A.F. Young, *"Superconductivity and strong correlations in moiré flat bands,"* Nature Physics **16**, 725-733 (2020).
[6] W. Pan, D.B. Burckel, C.D. Spataru, K.R. Sapkota, A.J. Muhowski, S.D. Hawkins, J.F. Klem, L.S. Smith, D.A. Temple, Z.A. Enderson, Z. Jiang, K. Thirunavukkuarasu, L. Xiang, M. Ozerov, D. Smirnov, C. Niu, P.D. Ye, P. Pai, and F. Zhang, *"Lithography-Defined Semiconductor Moirés with Anomalous In-Gap Quantum Hall States,"* Nano Letters **25**, 10536-10543 (2025).
[7] E. Suárez Morell, J.D. Correa, P. Vargas, M. Pacheco, and Z. Barticevic, *"Flat bands in slightly twisted bilayer graphene: Tight-binding calculations,"* Physical Review B **82**, 121407 (2010).
[8] R. Bistritzer, and A.H. MacDonald, *"Moiré bands in twisted double-layer graphene,"* Proceedings of the National Academy of Sciences **108**, 12233-12237 (2011).
[9] Y. Cao, V. Fatemi, S. Fang, K. Watanabe, T. Taniguchi, E. Kaxiras, and P. Jarillo-Herrero, *"Unconventional superconductivity in magic-angle graphene superlattices,"* Nature **556**, 43-50 (2018).
[10] Y. Cao, V. Fatemi, A. Demir, S. Fang, S.L. Tomarken, J.Y. Luo, J.D. Sanchez-Yamagishi, K. Watanabe, T. Taniguchi, E. Kaxiras, R.C. Ashoori, and P. Jarillo-Herrero, *"Correlated insulator behaviour at half-filling in magic-angle graphene superlattices,"* Nature **556**, 80-84 (2018).
[11] D. Waters, Y. Nie, F. Lüpke, Y. Pan, S. Fölsch, Y.C. Lin, B. Jariwala, K. Zhang, C. Wang, H. Lv, K. Cho, D. Xiao, J.A. Robinson, and R.M. Feenstra, *"Flat Bands and Mechanical Deformation Effects in the Moiré Superlattice of MoS(2)-WSe(2) Heterobilayers,"* ACS Nano **14**, 7564-7573 (2020).
[12] G. Abbas, Y. Li, H. Wang, W.-X. Zhang, C. Wang, and H. Zhang, *"Recent Advances in Twisted Structures of Flatland Materials and Crafting Moiré Superlattices,"* Advanced Functional Materials **30**, 2000878 (2020).
[13] H. Li, S. Li, M.H. Naik, J. Xie, X. Li, J. Wang, E. Regan, D. Wang, W. Zhao, S. Zhao, S. Kahn, K. Yumigeta, M. Blei, T. Taniguchi, K. Watanabe, S. Tongay, A. Zettl, S.G. Louie, F. Wang, and M.F. Crommie, *"Imaging moiré flat bands in three-dimensional reconstructed WSe2/WS2 superlattices,"* Nature Materials **20**, 945-950 (2021).
[14] S.K. Behura, A. Miranda, S. Nayak, K. Johnson, P. Das, and N.R. Pradhan, *"Moiré physics in twisted van der Waals heterostructures of 2D materials,"* Emergent Materials **4**, 813-826 (2021).
[15] C.N. Lau, M.W. Bockrath, K.F. Mak, and F. Zhang, *"Reproducibility in the fabrication and physics of moiré materials,"* Nature **602**, 41-50 (2022).
[16] Z. Dai, L. Liu, and Z. Zhang, *"Strain Engineering of 2D Materials: Issues and Opportunities at the Interface,"* Advanced Materials **31**, 1805417 (2019).
[17] N.P. Kazmierczak, M. Van Winkle, C. Ophus, K.C. Bustillo, S. Carr, H.G. Brown, J. Ciston, T. Taniguchi, K. Watanabe, and D.K. Bediako, *"Strain fields in twisted bilayer graphene,"* Nature Materials **20**, 956-963 (2021).
[18] Y. Hou, S. Zhang, Q. Li, L. Liu, X. Wu, and Z. Zhang, *"Evaluation local strain of twisted bilayer graphene via moiré pattern,"* Optics and Lasers in Engineering **152**, 106946 (2022).
[19] N.N.T. Nam, and M. Koshino, *"Lattice relaxation and energy band modulation in twisted bilayer graphene,"* Physical Review B **96**, 075311 (2017).
[20] H. Yoo, R. Engelke, S. Carr, S. Fang, K. Zhang, P. Cazeaux, S.H. Sung, R. Hovden, A.W. Tsen, T. Taniguchi, K. Watanabe, G.-C. Yi, M. Kim, M. Luskin, E.B. Tadmor, E. Kaxiras, and P.

Kim, *"Atomic and electronic reconstruction at the van der Waals interface in twisted bilayer graphene,"* Nature Materials **18**, 448-453 (2019).
[21] A. Weston, Y. Zou, V. Enaldiev, A. Summerfield, N. Clark, V. Zólyomi, A. Graham, C. Yelgel, S. Magorrian, M. Zhou, J. Zultak, D. Hopkinson, A. Barinov, T.H. Bointon, A. Kretinin, N.R. Wilson, P.H. Beton, V.I. Fal'ko, S.J. Haigh, and R. Gorbachev, *"Atomic reconstruction in twisted bilayers of transition metal dichalcogenides,"* Nature Nanotechnology **15**, 592-597 (2020).
[22] C. Liu, T.L. Hughes, X.-L. Qi, K. Wang, and S.-C. Zhang, *"Quantum Spin Hall Effect in Inverted Type-II Semiconductors,"* Physical Review Letters **100**, 236601 (2008).
[23] I. Knez, R.-R. Du, and G. Sullivan, *"Evidence for Helical Edge Modes in Inverted $\mathrm{InAs}/\mathrm{GaSb}$ Quantum Wells,"* Physical Review Letters **107**, 136603 (2011).
[24] W. Yu, V. Clericò, C.H. Fuentevilla, X. Shi, Y. Jiang, D. Saha, W.K. Lou, K. Chang, D.H. Huang, G. Gumbs, D. Smirnov, C.J. Stanton, Z. Jiang, V. Bellani, Y. Meziani, E. Diez, W. Pan, S.D. Hawkins, and J.F. Klem, *"Anomalously large resistance at the charge neutrality point in a zero-gap InAs/GaSb bilayer,"* New Journal of Physics **20**, 053062 (2018).
[25] J. Li, W. Yang, and K. Chang, *"Spin states in InAs/AlSb/GaSb semiconductor quantum wells,"* Physical Review B **80**, 035303 (2009).
[26] G.D. Sanders, Y. Sun, F.V. Kyrychenko, C.J. Stanton, G.A. Khodaparast, M.A. Zudov, J. Kono, Y.H. Matsuda, N. Miura, and H. Munekata, *"Electronic states and cyclotron resonance in n-type InMnAs,"* Physical Review B **68**, 165205 (2003).
[27] R.M. Wood, D. Saha, L.A. McCarthy, J.T. Tokarski, G.D. Sanders, P.L. Kuhns, S.A. McGill, A.P. Reyes, J.L. Reno, C.J. Stanton, and C.R. Bowers, *"Effects of strain and quantum confinement in optically pumped nuclear magnetic resonance in GaAs: Interpretation guided by spin-dependent band structure calculations,"* Physical Review B **90**, 155317 (2014).
[28] M.G. Burt, *"The justification for applying the effective-mass approximation to microstructures,"* Journal of Physics: Condensed Matter **4**, 6651 (1992).
[29] B.A. Foreman, *"Elimination of spurious solutions from eight-band k.p theory,"* Physical Review B **56**, R12748-R12751 (1997).
[30] B.A. Foreman, *"Analytical Envelope-Function Theory of Interface Band Mixing,"* Physical Review Letters **81**, 425-428 (1998).
[31] B.A. Foreman, *"Strong Linear- $k$ Valence-Band Mixing at Semiconductor Heterojunctions,"* Physical Review Letters **86**, 2641-2644 (2001).
[32] M. Willatzen, and L.C. Lew Yan Voon, The k p Method, 1st ed., Springer Berlin, Heidelberg, 2009.
[33] I. Vurgaftman, J.R. Meyer, and L.R. Ram-Mohan, *"Band parameters for III–V compound semiconductors and their alloys,"* Journal of Applied Physics **89**, 5815-5875 (2001).
[34] Y. Jiang, S. Thapa, G.D. Sanders, C.J. Stanton, Q. Zhang, J. Kono, W.K. Lou, K. Chang, S.D. Hawkins, J.F. Klem, W. Pan, D. Smirnov, and Z. Jiang, *"Probing the semiconductor to semimetal transition in InAs/GaSb double quantum wells by magneto-infrared spectroscopy,"* Physical Review B **95**, 045116 (2017).
[35] M.K. Hudait, M. Clavel, P.S. Goley, Y. Xie, J.J. Heremans, Y. Jiang, Z. Jiang, D. Smirnov, G.D. Sanders, and C.J. Stanton, *"Structural, morphological and magnetotransport properties of composite semiconducting and semimetallic InAs/GaSb superlattice structure,"* Materials Advances **1**, 1099-1112 (2020).
[36] Y. Jiang, M. Ermolaev, G. Kipshidze, S. Moon, M. Ozerov, D. Smirnov, Z. Jiang, and S. Suchalkin, *"Giant g-factors and fully spin-polarized states in metamorphic short-period InAsSb/InSb superlattices,"* Nature Communications **13**, 5960 (2022).
[37] Y. Jiang, M. Ermolaev, S. Moon, G. Kipshidze, G. Belenky, S. Svensson, M. Ozerov, D. Smirnov, Z. Jiang, and S. Suchalkin, *"$g$-factor engineering with InAsSb alloys toward zero band gap limit,"* Physical Review B **108**, L121201 (2023).

[38] R.L. Petritz, *"Theory of an Experiment for Measuring the Mobility and Density of Carriers in the Space-Charge Region of a Semiconductor Surface,"* Physical Review **110**, 1254-1262 (1958).
[39] W. Nakwaski, *"Effective masses of electrons and heavy holes in GaAs, InAs, A1As and their ternary compounds,"* Physica B: Condensed Matter **210**, 1-25 (1995).
[40] R. Knobel, N. Samarth, J.G.E. Harris, and D.D. Awschalom, *"Measurements of Landau-level crossings and extended states in magnetic two-dimensional electron gases,"* Physical Review B **65**, (2002-06-20).
[41] D. Shoenberg, Magnetic oscillations in metals, Cambridge university press, 1984.

Supplemental Material: Flat-Band Generation in InAs/GaSb Quantum Wells through Vertically Engineered Heterostructures

Zachery A. Enderson[1,2,3,*†], Jiyuan Fang[4,†], Wei-Chen Wang[4], Li Xiang[5], Mykhaylo Ozerov[5], Dmitry Smirnov[5], Zhigang Jiang[4,*], Samuel D. Hawkins[6], Aaron J. Muhowski[6], John F. Klem[6], and Wei Pan[2,*]

[1] Oak Ridge Institute for Science and Education IC Postdoctoral Fellowship 37831, USA
[2] Sandia National Laboratories, Livermore, CA 94551, USA
[3] Institute for Matter and Systems, Georgia Institute of Technology, Atlanta, GA 30332, USA
[4] School of Physics, Georgia Institute of Technology, Atlanta, GA 30332, USA
[5] National High Magnetic Field Laboratories, Tallahassee, FL 32310, USA
[6] Sandia National Laboratories, Albuquerque, NM 87185, USA

*Corresponding author. Email: zenderson3@gatech.edu (ZAE); zhigang.jiang@physics.gatech.edu (ZJ); wpan@sandia.gov (WP)
†These authors contributed equally to this work.

## Additional information for eight-band k·p calculations

Our k·p calculations, including zero-field band structure calculations, Landau level spectrum calculations, and optical absorption spectra calculations, closely follow our prior work [34], but are applied here to the quad-layer structure. The calculated zero-field band structure is shown in Fig. 2A; the calculated Landau level spectra are shown in supplemental Figs. S1,S2 below; the calculated optical absorption spectra are shown in Fig. 6B.

Landau fan diagram

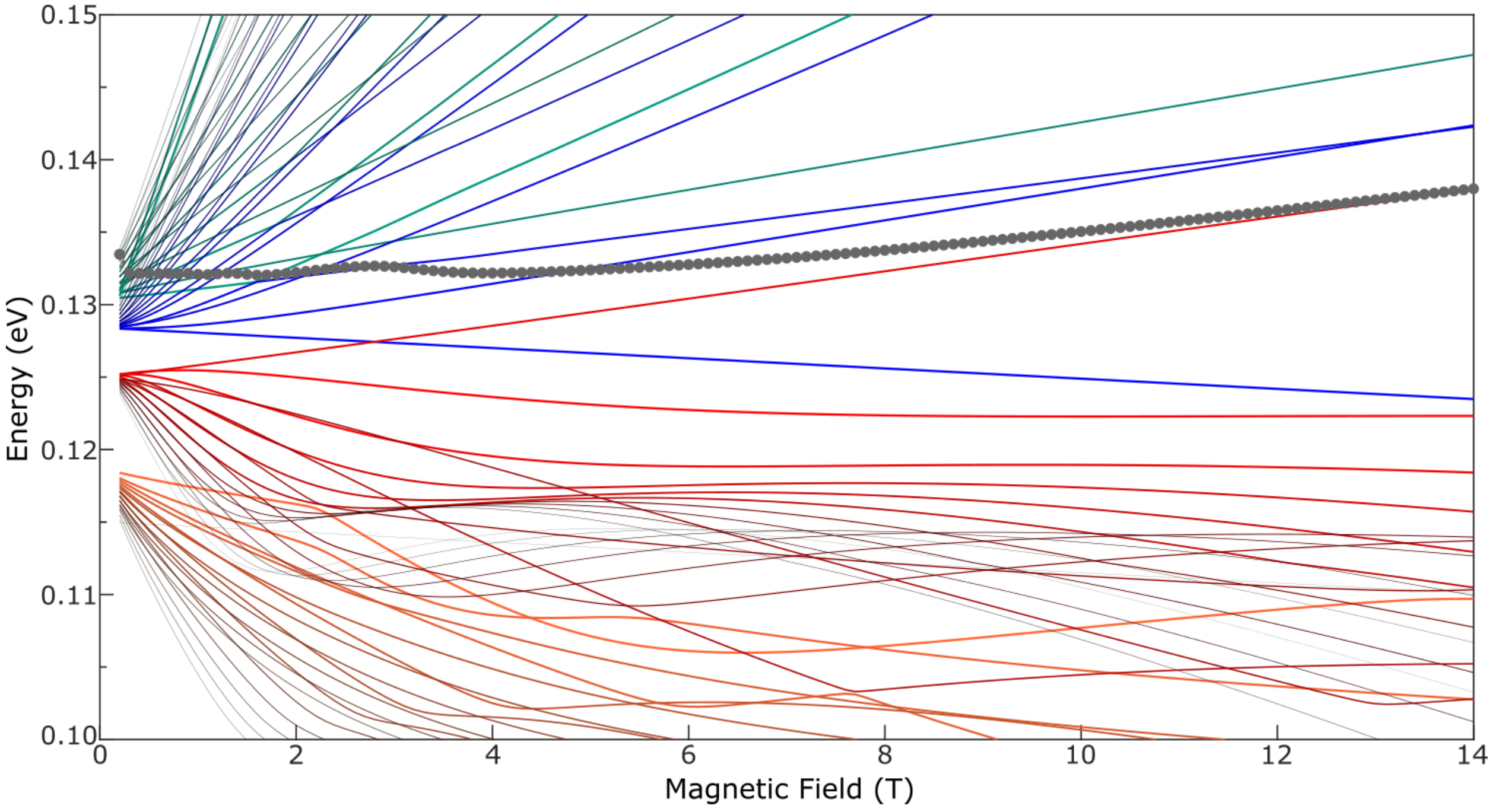

**Fig. S1. k·p calculated Landau fan diagram (full range).** k·p calculated Landau fan diagram over a 0.25–14 T magnetic field range. The plot shows the two highest valence bands (red hues) and the two lowest conduction bands (blue hues). The dotted gray line indicates the Fermi level.

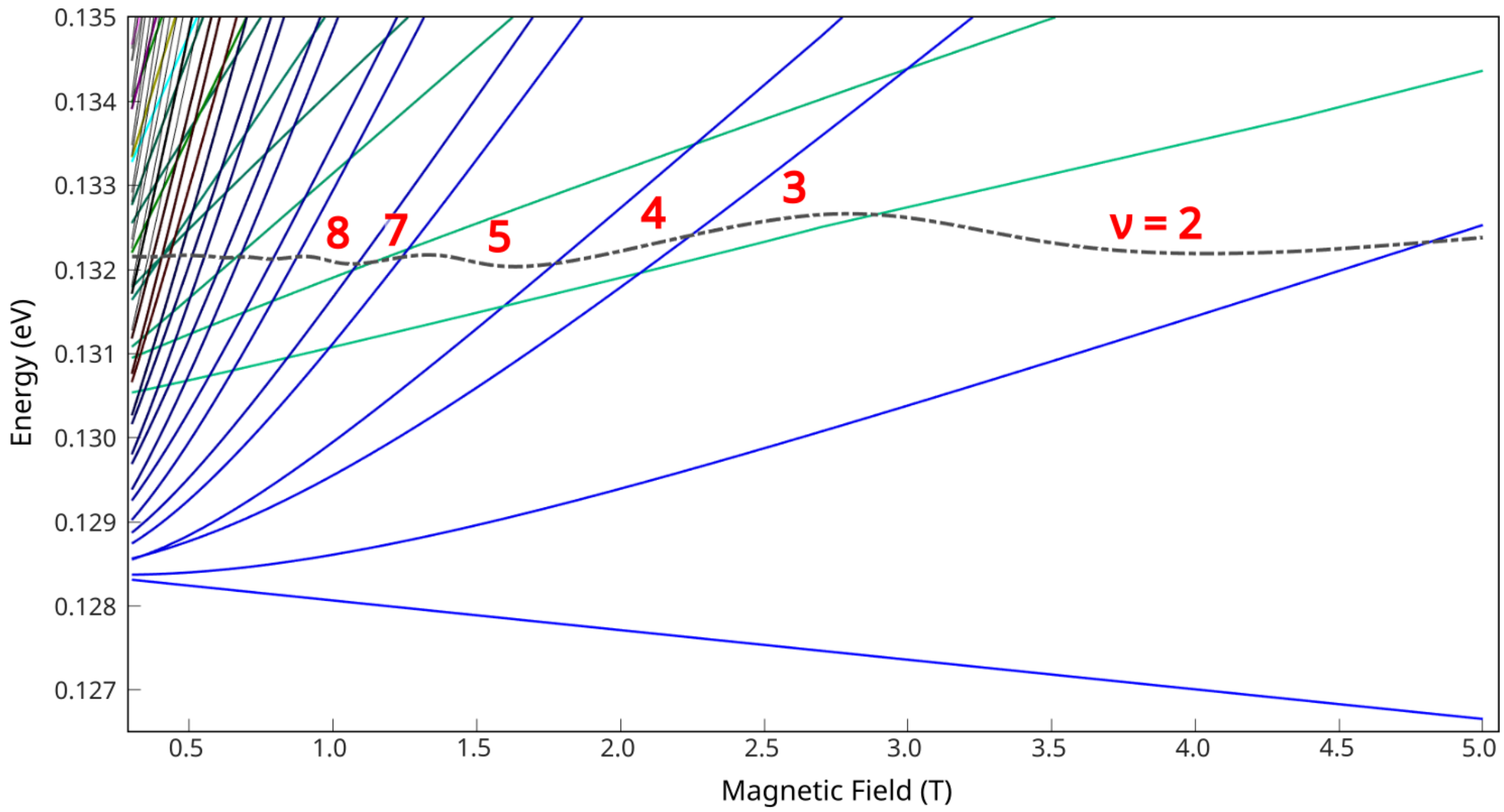

**Fig. S2. k·p calculated Landau fan diagram (reduced range).** k·p calculated Landau fan diagram over a 0.3–5 T magnetic field range. The plot shows the two lowest conduction bands (blue hues), and the dashed gray line indicates the Fermi level. Of note are the interband crossings that occur near the Fermi level. The corresponding Landau level filling factors are also labeled. The absence of SdH oscillations at filling factors $\nu = 6$ and $\nu = 11$ in the experiment is due to the Landau level crossings between the two conduction bands at $B \approx 1.25$ T and 0.62 T, respectively.

## Data analysis/processing

Magnetoresistance data was initially smoothed using a Savitzky-Golay (SG) filter with a smoothing window of $< 50$ mT. The data was converted from units of voltage (measured output from lock-in amplifier) to resistance with a 100 nA drive current. For Figs. 3 and 4, no further processing was performed.

For the temperature dependent Shubnikov-de Haas (SdH) oscillation data, a 2$^{nd}$-order polynomial background was fitted to a heavily SG filtered (smoothing window ~1/3 spectrum length) $R_{xx}$ curve and removed from the data described above. This data is shown in the upper inset of Fig. 5. The amplitude of the oscillation at a given temperature was calculated at each trough corresponding to an integer filling factor. This was done by finding the local maxima and local minima for the neighboring crests and trough, respectively, performing a linear fit to the two local maxima, using the fit to find the estimated amplitude at the $B$-field value corresponding to the trough minima, and subtracting the two resistance values (estimated amplitude and trough minima) to find $\Delta R_{xx}$. The process is illustrated in the lower inset of Fig. 5 for a given filling factor and temperature. It was repeated for all temperatures to give a data set shown in the main panel of Fig. 5. This data was fitted to the thermal dampening term in the Lifshitz-Kosevich (L-K) formula to calculate the value of $m^*$ (as outlined in the main text). The results for all relevant filling factors are shown in Table 1.

The far-infrared magneto-spectroscopy data presented in Fig. 6A was smoothed using an SG filter along the wavenumber axis at a constant magnetic field with a smoothing window $< 5$ cm$^{-1}$. A mean transmission spectrum taken from the full dataset ($T_{ave} = \int T(B)dB$) was subtracted from each individual spectrum taken at constant $B$, $T(B)/T_{ave}$. This normalized data is presented in Fig. 6A.

## Additional information for SdH oscillation analysis

<u>Temperature dependent background in magnetoresistance $R_{xx}$</u>

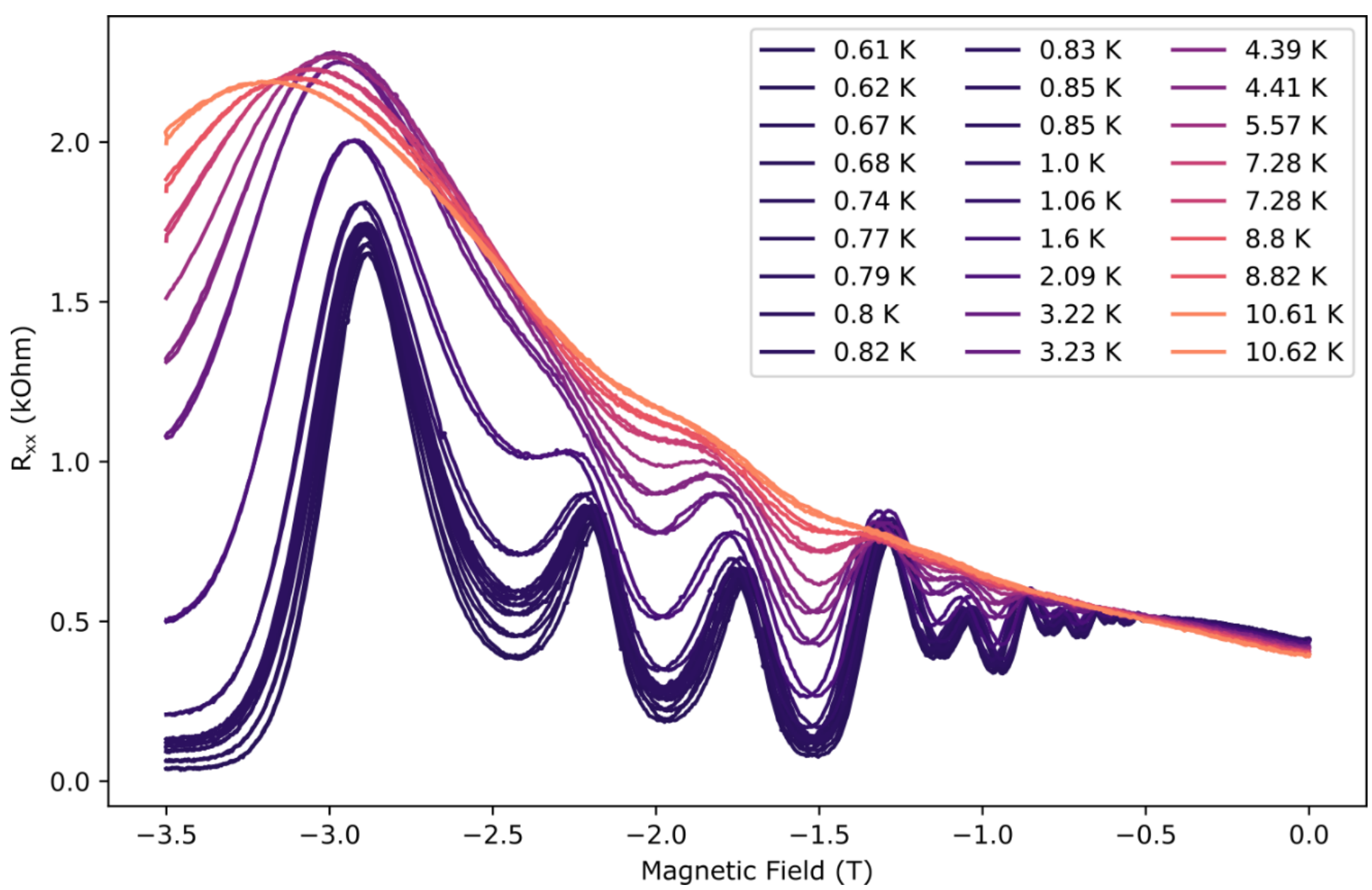

**Fig. S3. Original temperature dependent SdH oscillations.**

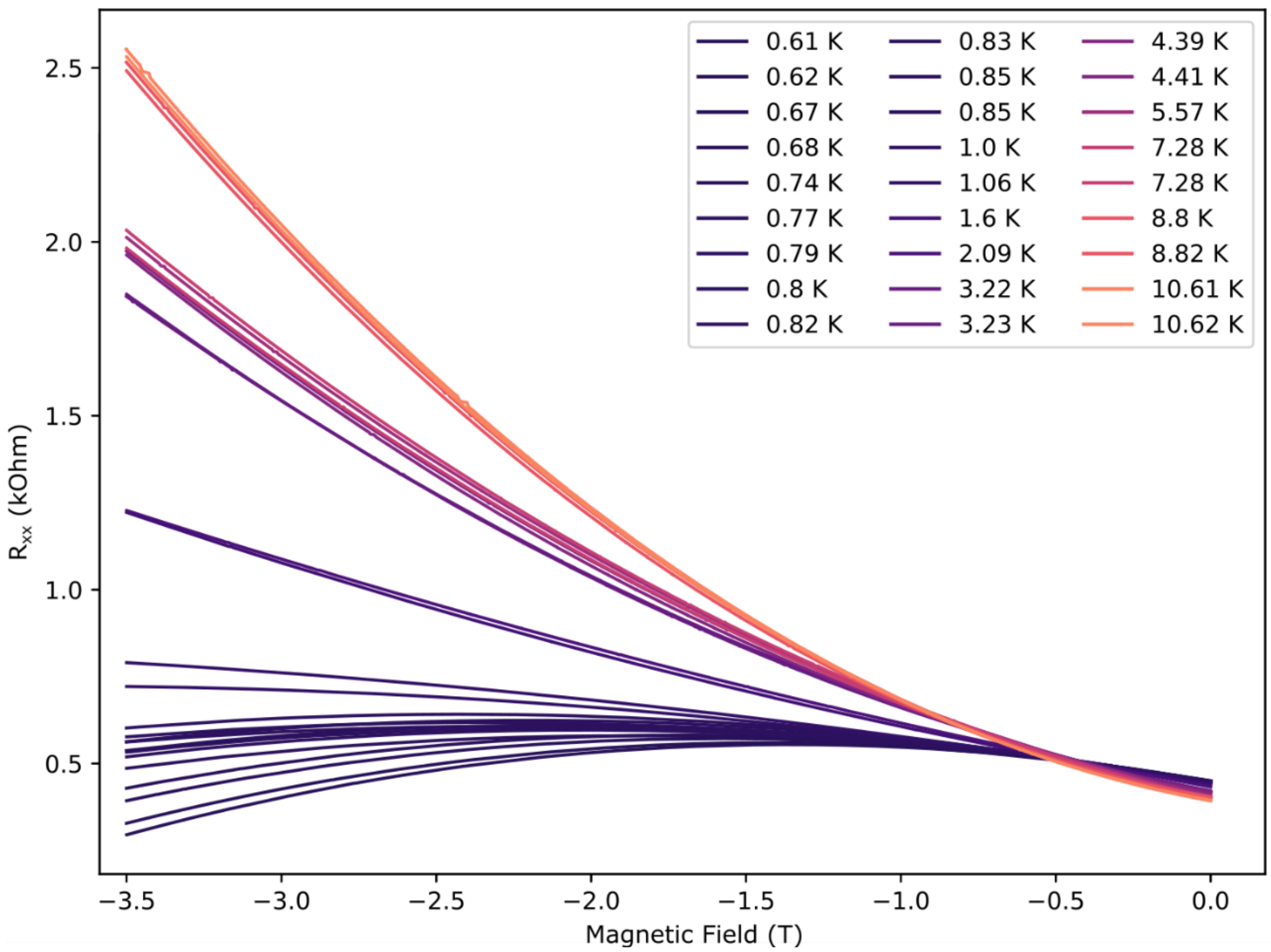

**Fig. S4. 2$^{nd}$-order polynomial background removed from original temperature dependent SdH oscillation data.**

<u>Fourier analysis of SdH oscillations</u>

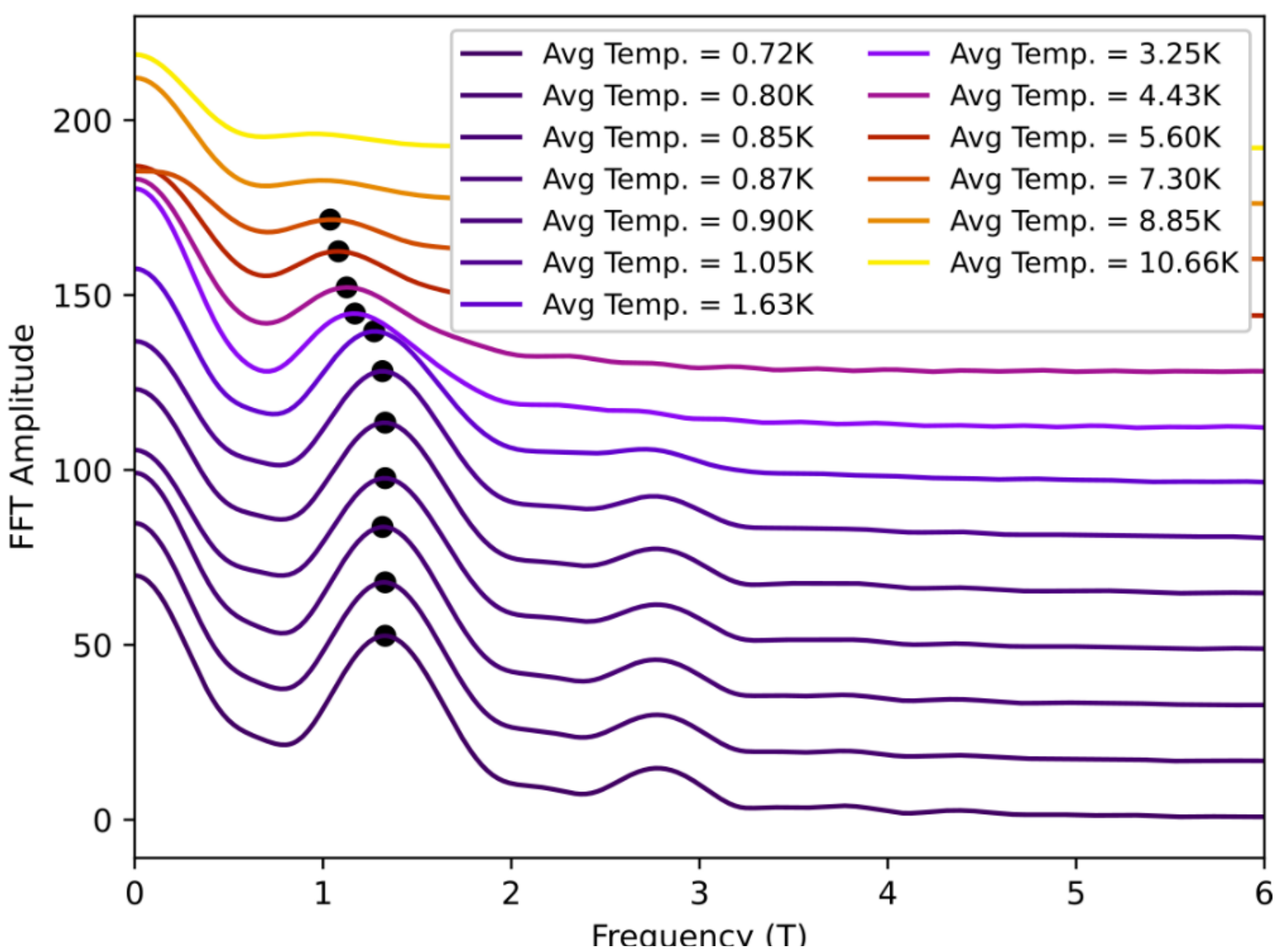

**Fig. S5. Fourier transform of temperature dependent SdH oscillations.** Fourier transform of the experimental SdH oscillation curves ($R_{xx}$) plotted as a function of $1/B$ at various temperatures with a 2nd-order polynomial background removed. The $R_{xx}$ data has been cropped to only include oscillations corresponding to $\nu > 3$. Traces are offset vertically for clarity.

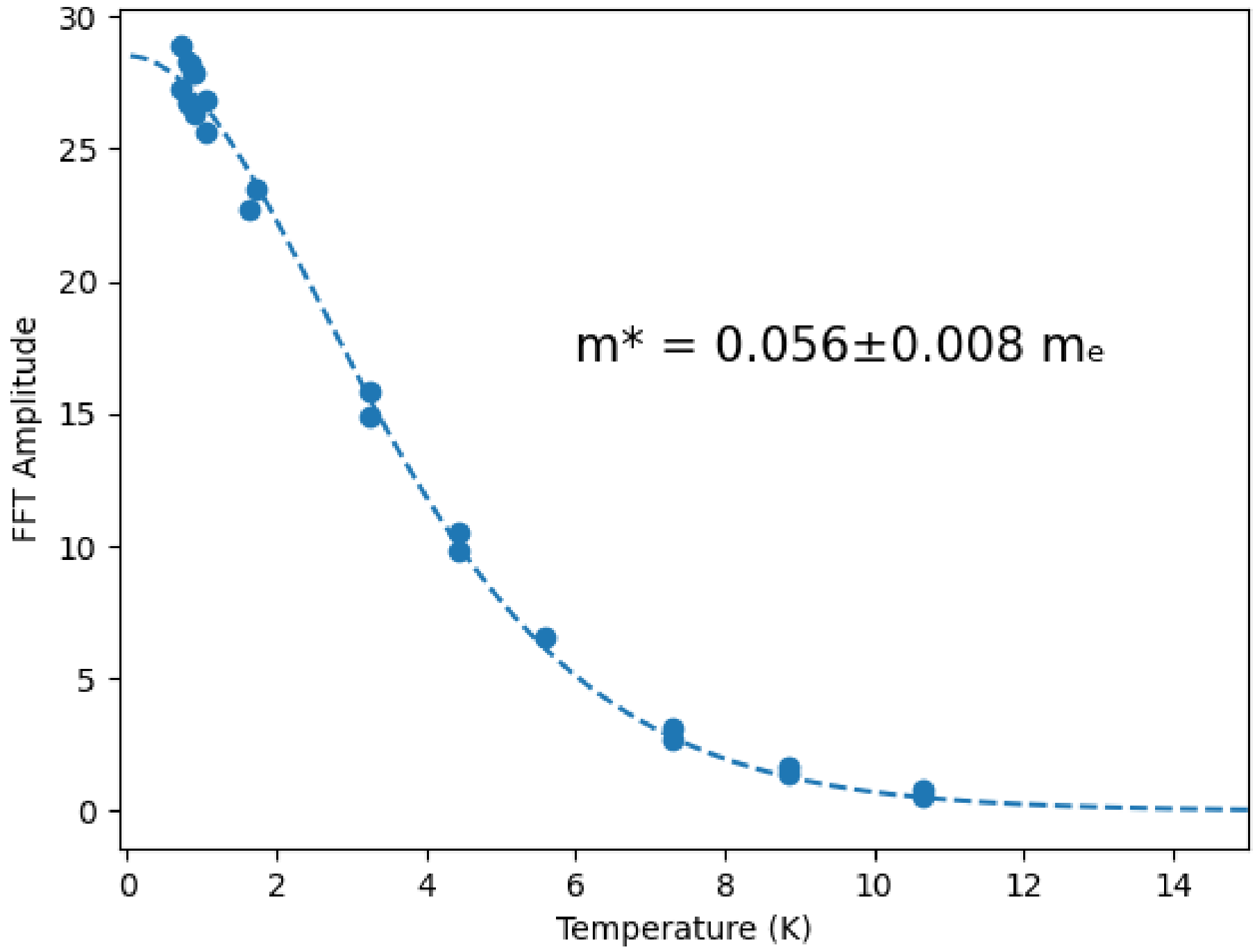

**Fig. S6. Effective mass deduced using Fourier analysis.** The amplitudes of the Fourier transform (indicated in Fig. S5) fit using the L-K formula to calculate the effective mass.